\documentclass[12pt]{article}
\RequirePackage{graphicx}

\newcommand{\commentout}[1]{}
\begin{document}
{\normalsize
\title{Magnetic susceptibility due to disorder-induced\\
neutral solitons in interacting polymer chains
\\[0.5cm]}
\author{Marc~Thilo~Figge\footnote{e-mail: figge@phys.rug.nl}, 
Maxim~Mostovoy, and
Jasper~Knoester\\
Institute for Theoretical Physics and
Materials Science Center\\
University of Groningen, Nijenborgh 4, 9747 AG Groningen,
The Netherlands
}
\date{\today}
\maketitle
\begin{abstract}
\normalsize
\baselineskip 28pt
%
%
%
We study the magnetic response due to neutral solitons induced by
disorder in polymer materials.  We account for interchain
interactions, which, if sufficiently strong, result in a
bond-ordered phase, in which the neutral solitons are bound into
pairs.  We analytically calculate the corresponding pair size
distribution.  As the spins of the solitons have a distance
dependent antiferromagnetic coupling, this allows us to calculate
the magnetic susceptibility in the ordered phase.  At low
temperatures, the result deviates from the usual Curie behavior
in a way that depends on the relative strength of the disorder
and the interchain interactions.  
We compare our results to the observed magnetic susceptibility 
of {\it trans}-polyacetylene and we suggest new experiments
extending towards lower temperatures.
\end{abstract}

\vspace{1.0cm}


\newpage
%
%

\section{Introduction}\label{intro}

%
%
In recent papers we have studied the effect of off-diagonal
disorder on the lattice configuration of a half-filled
Peierls-Hubbard chain with a doubly degenerate ground state
\cite{Mostovoy97,Mostovoy98}.  An example is the $\pi$-conjugated
polymer {\it trans}-polyacetylene, which in the absence of
disorder has a uniformly dimerized ground state.  As the energy
does not depend on the sign of the lattice dimerization (which
determines whether even or odd bonds are short), this ground
state is doubly degenerate.  For a single Peierls-Hubbard chain
we showed that arbitrarily weak off-diagonal disorder induces
neutral solitons (kinks) in the lattice dimerization, which
interpolate between the two degenerate bond alternations.  Even
though the creation energy of a neutral soliton is rather large
(of the order of the gap in {\it trans}-polyacetylene), the
energy loss is compensated by allowing the sign of the
dimerization to adjust to the electronic disorder fluctuations.
Off-diagonal disorder in conjugated polymers (i.e., disorder in
the hopping amplitudes of the $\pi$-electrons) may originate from
random chain twists, which decrease the overlap between the
$\pi$-orbitals of neighboring carbon atoms.  The density of
disorder-induced neutral solitons in a single chain is
proportional to the disorder strength
\cite{Mostovoy97,Mostovoy98}.

In {\it trans}-polyacetylene the disorder strength is presumably
rather large, as the average conjugation length in this material
is of the order of several tens of carbon atoms
only\cite{Silbey96}.  One would then expect the neutral solitons
to contribute significantly to the polymer's magnetic and optical
properties, as they carry spin $\frac{1}{2}$ and result in the
appearance of electronic states inside the Peierls gap.

There is, however, no direct evidence for the existence of a high
density of solitons in undoped {\it trans}-polyacetylene.
Electron spin resonance (ESR) experiments report only about one 
free spin per 3000 carbon atoms
\cite{Shirakawa78,Goldberg79,Weinberger80,Foot86}.  
Moreover, it appears difficult to observe neutral solitons in 
optical absorption experiments, as, contrary to what is expected 
from the Su-Schrieffer-Heeger (SSH) model \cite{Su79}, they seem 
not to give rise to a clear midgap absorption peak \cite{Weinberger84}.  
This may be explained by assuming that the on-site Coulomb repulsion 
$U$ is strong enough to shift the midgap peak towards the absorption 
edge resulting from interband transitions 
\cite{Campbell8687,Vardeny8586}.
Since both the peak and the absorption edge are significantly
broadened by the quantum lattice motion \cite{Mostovoy96}, there
may be no clear distinction between them.

X-ray scattering data do yield some indirect evidence for
disorder-induced kinks.  There still is considerable disagreement
whether neighboring carbon chains are dimerized in phase
($P2_{1}/a$ space group) or in anti-phase ($P2_{1}/n$ space
group) \cite{Martens94,Bott87}.  While this may originate from
different preparation methods leading to different space groups,
it has also been pointed out that the disagreement may result
from a high density (of the order of several percent) of kinks
that locally change the relative sign of the dimerization in
neighboring chains \cite{Kahlert87,Perego88}.  Yet, it is not
clear how such random changes would lead to sharp peaks in the
x-ray spectra.

From the above it appears that a clear signature of the effect of
disorder-induced solitons is still to be found.  This has
motivated us to study the magnetic response of disorder-induced
solitons in more detail.  A proper modeling of the magnetic 
susceptibility involves more than a calculation of the density of 
solitons in an isolated chain.  As we noted in
Refs.~\cite{Mostovoy97,Mostovoy98}, the actual density of neutral
solitons (and thus of spins) is determined by the competition
between disorder and interchain interactions, as the latter lead
to confinement of soliton-antisoliton pairs and may restore the
long-range bond order.  Moreover, at low temperature, the
exchange between the spins of neighboring solitons on a single
chain tends to bind them into a singlet state\cite{Lin-Liu80},
which has no magnetic response.  Thus, the magnetic
susceptibility and, in particular, its temperature dependence
should be expected to depend strongly on the interplay between
disorder and interchain interactions.

In this paper, we focus on the magnetic susceptibility of
disorder-induced solitons in the phase with long-range bond
order.  In this case, solitons occur in isolated pairs of random
size dictated by the disorder realization.  We study the
statistics of these pairs by mapping the problem on the
anisotropic random-field Ising model, which is treated in the
chain-mean-field approximation \cite{Figge98}.  This mapping is
analogous to what we did in Refs.~\cite{Mostovoy97,Mostovoy98}
for isolated chains.  We briefly explain the mapping in
Sec.~\ref{model} and discuss the phase diagram of this model.  In
Sec.~\ref{masu}, we express the magnetic susceptibility of an
ensemble of soliton-antisoliton pairs in terms of the, as yet unknown,
distribution of exchange interactions.  In Sec.~\ref{psd}, we
calculate the distribution of soliton-pair sizes using the
saddle-point method. 
From the pair size distribution we derive in Sec.~\ref{conti}
the distribution of exchange constants.
The latter is used in Sec.~\ref{appl} to calculate the magnetic
susceptibility of {\it trans}-polyacetylene. 
We find that the low-temperature behavior of this susceptibility
deviates from the Curie law and we fit our results to the 
experimental data obtained in Ref.~\cite{Foot86}.
In Sec.~\ref{sumconcl}, we summarize and conclude.
%
%

\section{Solitons in Interacting Disordered Peierls Chains}
\label{model}

%
%
In Ref.\cite{Mostovoy98}, we have shown that the statistics of
neutral solitons in isolated weakly disordered Peierls chains can
be studied using the one-dimensional random-field Ising model
(RFIM).  
In this mapping, Ising variables $\sigma_m = \pm 1$ ($m=1, ..., M$)
are defined on the sites of a lattice with lattice constant
$d$.
These variables play the role of the sign of the dimerization in 
the Peierls chain, while the random "magnetic" field $h_m$ at site 
$m$ represents the off-diagonal disorder, which locally lifts the 
degeneracy between the two dimerization phases in the Peierls chain.  
Two neighboring sites on the lattice having different Ising variable,
correspond to the occurrence of a soliton in the Peierls chain.
Therefore, the creation energy $\mu$ of a soliton in the Peierls
chain is equivalent to the exchange interaction between
neighboring Ising spins.  
In the SSH model of {\it trans}-polyacetylene 
$\mu=2\Delta_0/\pi\simeq 0.5$ eV ($\Delta_0$ is the dimerization).
We emphasize, however, that this mapping is not limited to the
SSH model, but also holds in the presence of electron-electron
interactions, in which case the value of $\mu$ is smaller 
\cite{Kivelson82,Campbell84}.

Our approach may easily be extended to account for
three-dimensional effects: interchain interactions (electron
hopping, eleastic forces, or Coulomb interactions) tend to favor
a coherence of the dimerization pattern on neighboring chains,
which in Ising language translates into an interaction, $2W$,
between spins on neighboring chains.  As for
quasi-one-dimensional materials, like conjugated polymers, $W \ll
\mu$, we are thus dealing with a strongly anisotropic
random-field Ising model \cite{Figge98}.  The anisotropy allows
one to treat the interchain interactions in a mean-field way, an
approach known as the chain-mean-field approximation.  The energy
of the resulting Ising model is given by:
\begin{equation}
E\{\sigma_{m}\}\;=\sum_{m=1}^{M} \left[
\frac{\mu}{2} (1-\sigma_{m}\sigma_{m+1})\;
-\;h_{m}\sigma_{m}\;
-\;B\sigma_{m}\,\right]\;,
\label{rfim}
\end{equation}
Here, the first term describes the energy cost for creating
$\frac{1}{2} \sum_{m=1}^{M} (1-\sigma_{m}\sigma_{m+1})$ kinks and
the second term describes the interaction energy with the random
magnetic field.  The latter is assumed to have a Gaussian
distribution with zero mean ($\langle h_m \rangle\;=\;0$) and
correlator:
\begin{equation}
\langle h_m h_n \rangle\;=\;\epsilon\,\delta_{m,n}\;,
\end{equation}
where $\epsilon$ is the disorder strength.  
We consider the case of weak disorder: $\epsilon\ll \mu^2$. 
Finally, the third term in Eq.~(\ref{rfim}) describes the interchain
interactions in the mean-field approximation, where the
homogeneous ``magnetic'' field $B$ is proportional to the average
order parameter:
\begin{equation}
B\;=\;
W\,\langle\langle\sigma\rangle\rangle\;.
\label{Bdef}
\end{equation}
The double brackets denote both the thermal and the random-field
average.

To end the explanation of our model, a few remarks are in
place.  First, by replacing the dimerization by a discrete Ising
variable, we have neglected the true dimerization profile 
associated with a soliton.  As the extent of this profile is
given by the correlation length $\xi_0$, our "sudden-kink
approximation" is valid as long as the soliton density is small
compared to $1/\xi_0$, as is the case for weak disorder.    
Taking into account the true dimerization profile results in an 
effective increase of the kink creation energy of the order 
$\sqrt{\epsilon\, \xi_0/a}$ (with $a$ the average carbon-carbon distance 
in {\it trans}-polyacetylene) and, thus, in a small reduction of
the soliton density. 
This has recently been confirmed explicitely in numerical 
simulations which do account for the true profile 
\cite{Mertsching98}.
Second, it should be kept in mind that
the RFIM Eq.~(\ref{rfim}) is an effective model, obtained by
integrating out small lattice fluctuations.  As a result, the
kink creation energy $\mu$ weakly depends on the temperature 
\cite{Mostovoy98}.  
Third, above we have not specified the value of the Ising lattice
constant $d$ (which should not be confused with the lattice
constant $a$ of the polymer chain). 
It should be noted that both the disorder strength $\epsilon$ 
and the interchain interaction energy $W$ scale proportional to 
$d$.
In Sec.~\ref{conti} we show that all physical observables are
$d$-independent in the $d\rightarrow 0$ limit.

The temperature versus disorder strength phase diagram of the
model Eq.~(\ref{rfim}) contains two phases: the ordered phase
characterized by a nonzero value of the average dimerization,
$\langle\langle \sigma \rangle\rangle \Delta_0$, and the
disordered phase, in which the long-range bond order (LRBO) is
destroyed by thermal and disorder-induced kinks.  The two
phases are separated by a second-order transition.  Figure 1
shows the phase diagram calculated for $W/\mu=0.008$, a typical
value for {\it trans}-polyacetylene if the interchain
interactions are dominated by interchain electron hopping
\cite{Baeriswyl83}.  
The stars in Fig.~1 denote the phase boundary which we
obtained by numerical simulation of the model Eq.~(\ref{rfim})
using an algorithm based on the transfer-matrix approach (cf.
Ref.\cite{Mostovoy98}).  The order parameter $\langle\langle
\sigma \rangle\rangle$ was found from a self-consistent
calculation of the mean field $B$ and the critical curve was then
obtained by requiring that
$\langle\langle\sigma\rangle\rangle\rightarrow 0$.  The smooth
temperature dependence of the phase boundary was obtained by
averaging the free energy over $10^4$ random-field realizations
for a chain with $10^3$ sites.  The solid curve in Fig.~1
indicates the phase boundary which was calculated in
Ref.\cite{Figge98} from an analytical expression for the average
free energy of the continuum version of the model
Eq.~(\ref{rfim}).  With the exception of a small temperature
region, $T < T_0(\epsilon)$ \cite{Figge98}, the results of the
continuum and the discrete models agree well.

At low temperatures the phase transition results from a
competition between the disorder and the interchain interactions.
In fact, the critical disorder strength which separates the
phases with and without LRBO at zero temperature reads
\cite{Figge98}
\begin{equation}
\epsilon_c\approx \frac{2}{3}W\mu\;.
\label{epsilon_c}
\end{equation}

In the disordered phase, $\langle\langle\sigma\rangle\rangle=0$,
the density of disorder-induced neutral solitons (spin-flips) is to 
lowest order in  $\epsilon$ given by \cite{Figge98}
\begin{equation}
n_s\;=\;\frac{1}{d}\,\frac{\epsilon}{\mu^2}\;,
\end{equation}
as is the case for a single disordered chain ($W=0$)
\cite{Imry75b}. 
On the other hand, for $\epsilon < \epsilon_c$ the order parameter 
$\langle\langle\sigma\rangle\rangle$ is observed to increase rapidly 
\cite{Figge98} with a slope that is proportional to the ratio 
$\mu/W \gg 1$.
Thus, in the overwhelming part of the ordered phase the system
is nearly perfectly ordered with an order parameter close to unity, 
$\langle\langle\sigma\rangle\rangle\simeq 1$, and the solitons are bound 
into pairs by the interchain interactions. 
Well within the ordered phase ($\epsilon\ll\epsilon_c$) their density is
exponentially suppressed.  
The distance between the soliton-antisoliton pairs is much larger than 
the typical pair size and the number of soliton-antisoliton pairs per 
unit length reads \cite{Figge98}
\begin{equation}
n_p\;=\;\frac{1}{d}\,\frac{2\, W^2}{\epsilon}\,
\exp\left(-\,2\,\frac{W\,\mu}{\epsilon}\right)\;
\label{contidens}
\end{equation}
(for $W^2 \ll \epsilon \ll 2W \mu /3$).

In the following we will focus on the LRBO phase and calculate
the magnetic susceptibility due to the spins of the bound pairs
of neutral solitons.
%
%

\section{Magnetic susceptibility in the ordered phase}
\label{masu}

%
%
Apart from the interchain interaction discussed in the previous
section, there is also an intrachain interaction between kinks.
The latter interaction is strong only when the distance between
kinks is of the order of their size, $\xi_0$.  Thus, for weak
disorder, when the density of kinks is small, it has little
effect on the statistics of the kinks.  It may, however, be
important for the magnetic properties of disordered Peierls
systems, as it results in an antiferromagnetic exchange between
the spins of neutral kinks \cite{Lin-Liu80}.  This exchange can
bind the spins of neighboring kinks into nonmagnetic singlets,
thus reducing the magnetic susceptibility of the system.

As argued in the previous section, for a nearly perfectly ordered
system the typical distance between disorder-induced
soliton-antisoliton pairs is much larger than the typical pair
size.  We may then neglect the spin exchange between kinks from
different pairs.  The Hamiltonian describing the interactions of
soliton and antisoliton spins, $\vec{S}_1$ and $\vec{S}_2$,
within one pair reads
\begin{equation}
\hat{H}_{pair}\;=\; J \left(\vec{S}_1 \cdot \vec{S}_2 -
\frac{1}{4}\right) - g\mu_B H \left(S_1^z + S_2^z\right),
\label{pairham}
\end{equation}
where $J$ is the exchange constant in the pair and $H$ is the
external magnetic field.  The pair free energy is given by
\begin{equation}
f(J, H)\;=\;-\,J\,-\,\frac{1}{\beta}\,\ln\left[
1\,+\,e^{-\beta J}\,\left(1\,+\,2\,
\cosh(\beta g \mu_B H)\right)
\right]
\label{pairfree}
\end{equation}
and the zero-field magnetic susceptibility of the pair is
\begin{equation}
\chi(T,J)\;=\;-\,
\frac{\partial^2 f(J, H)}{\partial H^2}\Bigg|_{H=0}=\;
2 g^2\mu_B^{2}\,
\beta\,\frac{e^{-\beta J}}{1\,+\,3 e^{-\beta J}}\;.
\label{pairsus}
\end{equation}

The coupling $J$ decreases with the soliton-antisoliton
separation $R$.  Quite generally, the large-$R$ behavior is
\begin{equation}
J\;=\;J_0\,\exp\left(-\frac{R}{\rho}\right)\;,
\label{JofRandrho}
\end{equation}
where $\rho=\xi_0/d$ ($R$ is measured in units of $d$) and $J_0$
is of the order of the spin gap.  For the SSH model, in which
Coulomb interactions are neglected and the spin gap equals the
charge gap, $J_0 = 4 \Delta_0$ \cite{Lin-Liu80}.

As $R$ is a random quantity that is imposed by the disorder
realization, also $J$ is random.  If we know the pair size
distribution, $p(R)$, the distribution of exchange values,
$w(J)$, can be obtained using Eq.~(\ref{JofRandrho}).  
We normalize the latter to the total density of spin-pairs
\begin{equation}
n_p\;=\;\int_{0}^{\infty}\! dJ\,w(J)\;.
\label{wnorm}
\end{equation}
The system's magnetic susceptibility is then given by
\begin{equation}
\chi(T)\;=\;\int_{0}^{\infty}\! dJ\,w(J)\,\chi(T,J)\;,
\label{chires}
\end{equation}
with $\chi(T,J)$ as in Eq.~(\ref{pairsus}).

Clearly, the temperature dependence of the magnetic susceptibility
is determined by the pair size distribution. As we will show
in detail in Secs.~\ref{psd} and \ref{conti}, in the LRBO phase, 
$p(R)$ is sharply peaked at some $R^{\ast}$, while for $R\gg R^{\ast}$
\begin{equation}
p(R\gg R^{\ast})\;\propto\;\exp\left(-\,\alpha\,\frac{R}{\rho}\right)\;
\label{pexpogen}
\end{equation}
with $\alpha$ a constant determined by the strength of the disorder
and interchain interactions.
Equation (\ref{pexpogen}) in a straightforward way yields a power-law
exchange distribution
\begin{equation}
w(J)\;\propto\;
\left(\frac{J_0}{J}\right)^{1-\alpha}\;
\label{powerlawex}
\end{equation}
for $J\ll J(R^{\ast})$. 
This part of $w(J)$ dictates the behavior of the magnetic
susceptibility at low temperature, $T\ll J(R^{\ast})$.
Pairs with $J\gg J(R^{\ast})$ (or: $R\ll R^{\ast}$) are
in the nonmagnetic singlet state at these low temperatures.
We thus find 
\begin{equation}
\chi(T\ll J(R^{\ast}))\;\propto\;\left(\frac{J_0}{T}\right)^{1-\alpha}\;,
\end{equation}
which deviates from the high temperature Curie behaviour.

We note that, in order to describe the anomalous temperature
dependence of the magnetic susceptibility of charge transfer
salts, Clark {\it et al.} \cite{Clark78} also introduced pairs
of spins with a random antiferromagnetic coupling.
These pairs, however, were introduced in a purely phenomenological
way, whereas in our model they naturally emerge as disorder-induced
soltion-antisoliton pairs with a distribution of exchange
constants that follows from the pair size distribution.

%
%
%
%
\section{Calculation of the Pair Size Distribution} \label{psd} 
%
%
%
The pair size distribution $p(R)$ is defined as the number of
soliton-antisoliton pairs of size $R$ per site of the Ising chain.  
For a given disorder realization $\{h_m\}$ one finds from
Eq.~(\ref{rfim}) that the energy of a configuration $(m_1, m_2)$,
with the soliton located between $m_1$ and $m_1+1$ and the antisoliton 
between $m_2$ and $m_2+1$, reads:
\begin{equation}
E[m_1,m_2]\;=\;E_0\,-\,\Delta
E[m_1,m_2]\;, \label{enbal}
\end{equation}
where $E_0$ denotes the energy for a configuration without
solitons and
\begin{equation}
\Delta E[m_1,m_2]\;=\;-2\mu\,-\,
2\!\!\sum_{m=m_1+1}^{m_2}\!\!(B+h_m)
\label{energybalance}
\end{equation}
is the energy change due to the creation of the soliton-antisoliton pair.  As
we restrict ourselves to isolated soliton-antisoliton pairs, it is sufficient
to consider a segment of the chain which contains one such pair
located far away from its end points.  Furthermore, because the
sequence of solitons and antisolitons along the chain is
determined by fixed boundary conditions for the lattice
dimerization, we may, without loss of generality, assume that
$m_2 > m_1$.  Then, the pair size $R$ (in units of the Ising
lattice constant $d$) is given by
\begin{equation}
R\;=\;m_2\,-\,m_1\;.
\label{defR}
\end{equation}

The soliton-antisoliton pair configuration $(m_1, m_2)$ is only energetically
favorable if
\begin{equation}
\Delta E[m_1,m_2]\, \geq\, 0\;.
\label{necesscond}
\end{equation}
This is, however, not sufficient to calculate the pair size
distribution $p(R)$, as we also have to impose the condition that
this pair configuration has lower energy than any other pair in
the considered chain segment. Thus, simultaneously, the energy of
the pair configuration has to satisfy the inequalities
\begin{equation}
\Delta E[m_1,m_2]\,\geq\,\Delta E[m_1^{\prime},m_2^{\prime}]\;,
\label{optimalcond}
\end{equation}
for all other possible pair configurations $(m_1^{\prime},m_2^{\prime})$.
Therefore, the desired pair size distribution takes the form:
\begin{equation}
p(R)\;=\;
\left\langle
\Theta\left(\,\Delta E[m_1,m_2]\,\right)\,
\prod_{(m_1^\prime,m_2^\prime)
}\,
\Theta\left(\,
\Delta E[m_1,m_2]-\Delta E[m_1^{\prime},m_2^{\prime}]
\,\right)
\right\rangle\;,
\label{pofR}
\end{equation}
%
where $\Theta(x)$ is the step function:
\begin{equation}
\Theta(x)\;=\;\left\{
\begin{array}{l@{\quad \quad}l}
1 & {\rm for}\;\; x \geq 0 \\
0 & {\rm for}\;\;x < 0
\nonumber
\end{array}\right.
\label{theta}
\end{equation}
and the brackets, $\langle ... \rangle$, denote the Gaussian
average over disorder realizations $\{h_m\}$.
The definition Eq.~(\ref{pofR}) ensures that, in accordance with 
Eq.~(\ref{wnorm}), $p(R)/d$ is normalized to the density of 
soliton-antisoliton pairs:
\begin{equation}
n_p\;=\;\frac{1}{d}\,\int^{\infty}_0\!\!dR\,p(R)\;.
\label{denspofR}
\end{equation}

It is easy to see that $p(R)$ factorizes into two independent
parts, $p_{out}$ and $p_{in}$, that account for the
soliton-antisoliton pairs with a size that is, respectively, larger and
smaller than $R$:
\begin{equation}
p(R)\;=\;
p_{out}\:p_{in}\;,
\label{pfactorize}
\end{equation}
with
\begin{eqnarray}
&&p_{out}\;=\;
\bigg\langle\;
\Theta(z+s_{m_1})
\;\Theta(2z+s_{m_1}+s_{m_1-1})
\;\Theta(3z+s_{m_1}+
s_{m_1-1}+s_{m_1-2})
\;...\;
\bigg\rangle\nonumber\\
&&\bigg\langle\;
\Theta(z+s_{m_2+1})
\;\Theta(2z+s_{m_2+1}+
s_{m_2+2})
\;\Theta(3z+s_{m_2+1}+
s_{m_2+2}+s_{m_2+3})
\;...\;
\bigg\rangle\;,\;\;\;\label{pout}
\end{eqnarray}
and
\begin{equation}
p_{in}\;=\;\bigg\langle\;
\Theta(- I - Rz -
\!\!\sum_{m=m_1+1}^{m_2}\!\!s_m)\label{pin}\;\Pi_L\;\Pi_R\;\bigg\rangle\;.
\end{equation}
Here, we have defined the dimensionless variables
$s_m=h_m/\sqrt{\epsilon}$, $z = B/\sqrt{\epsilon}$, $I =
\mu/\sqrt{\epsilon}$, while
\begin{equation}
\Pi_L\;\equiv\;\Theta(-z-s_{m_{1}+1})\;\Theta(-2z-s_{m_{1}+1}-s_{m_{1}+2})\;
...\; \Theta(-Rz-s_{m_{1}+1}-...-s_{m_{2}})
\end{equation}
and
\begin{equation}
\Pi_R\;\equiv\;\Theta(-z-s_{m_{2}})\;\Theta(-2z-s_{m_{2}}-s_{m_{2}-1})\;...\;
\Theta(-Rz-s_{m_{2}}-...-s_{m_{1}+1})\;.
\end{equation}

Note that $p_{out}$ itself also consists of two independent factors:
the first factor excludes the pairs with the soliton located to
the left of $m_1$, while the second one excludes antisoliton
positions larger than $m_2+1$. Both these factors can be written in
the form:
\begin{equation}
Y(z)\;=\;
\prod_{m=1}^{\infty}\left[
\int_{-\infty}^{+\infty}\!\!ds_m\,
f(s_m)\,
\Theta\Big(\sum_{k=1}^{m}(z+s_k)\Big)
\right]\;,
\label{yzla}
\end{equation}
where
\begin{equation}
f(s)=\frac{\exp(-\frac{1}{2}s^2)}{\sqrt{2\pi}}
\label{f}
\end{equation}
is the Gaussian weight. As a result, for the outer factor we obtain:
\begin{equation}
p_{out}\;=\;\left[Y(z)\right]^2\;.
\label{pout1}
\end{equation}
The function $Y(z)$ will be calculated later in this section.

The calculation of the inner factor is complicated by the
presence of the extra $\Theta$-function in Eq.~(\ref{pin}), which
precludes the factorization of $p_{in}$ in two independent
averages.  
However, considerable simplification is possible, because we focus
on the bond-ordered phase where $\langle\langle\sigma\rangle\rangle\simeq 1$.
Then the density of disorder-induced soliton-antisoliton pairs is small, 
and the main suppression factor in $p(R)$ is the probability of the disorder 
realization necessary to create a pair \cite{Ltail,Fluct}.  In other words,
the most important contribution to $p_{in}$ [and $p(R)$] comes
from averaging the first $\Theta$-function in Eq.~(\ref{pin}):
\begin{equation}
\bigg\langle
\Theta(-I\,-\!\!\sum_{m=m_1+1}^{m_2}\!\!(z+s_m))
\bigg\rangle\;=\;\frac{1}{2}\,{\rm erfc}(g(R))
\;\approx\;\frac{\exp(-g(R)^2)}{\sqrt{4\pi}\:g(R)}\;,
\label{justfluc}
\end{equation}
where
\begin{equation}
g(R)\;\equiv\;\frac{I+Rz}{\sqrt{2R}}\;.
\label{defofg}
\end{equation}
Here, the asymptotic expression for the complementary error function 
${\rm erfc}(g(R))$ was used because the minimal value for its argument is 
easily shown to be $g_{min}=\sqrt{3\epsilon_c / \epsilon}$, so that for 
$\epsilon \leq \epsilon_c/2$ the relative error becomes already less than 
several percent. 

The interpretation of this result is that the optimal disorder
fluctuation (i.e., the disorder realization with the largest
weight) that can induce a soliton-antisoliton pair of size $R$
has a constant value $-h_R$ in the interval of length $R$ and is
zero outside the interval.  The amplitude $h_R$ is determined
by the energy balance [see Eqs.~(\ref{energybalance}) and
(\ref{necesscond})]:
\begin{equation}
h_R R = \mu + W R\;.
\label{ebal}
\end{equation}
The weight of the optimal fluctuation,
$w = \exp\left(- R {h_R}^2/(2 \epsilon) \right)$,
is precisely the exponential factor appearing in Eq.~(\ref{justfluc}).
At
\begin{equation}
R^{\ast}\;=\;\frac{I}{z}\;=\;\frac{\mu}{W}\;\gg\;1
\label{estpeak}
\end{equation}
the weight reaches its maximal value,
$\exp\left(-2W\mu/\epsilon\right)$. For $\epsilon \ll
\epsilon_c$, the maximal weight is small [as was also found
in Eqs. (\ref{justfluc}) and (\ref{defofg})] and the
soliton-antisoliton pairs are suppressed. In that case, all
disorder realizations that contribute significantly to $p(R)$ are
close to the optimal fluctuation.

Bearing this in mind, we now calculate the inner factor
Eq.~(\ref{pin}).  First, we can rewrite Eq.~(\ref{pin}) in the form:
\begin{equation}
p_{in} = \int_{zR+I}^{\infty} dS
\int_{-i\infty}^{+i\infty} \!\!\frac{d\lambda}{2\pi i}
e^{-\lambda S}\prod_{m=m_1+1}^{m_2}
\int_{-\infty}^{+\infty}\!\!ds_m e^{-\lambda s_m}f(s_m) \Pi_L \Pi_R\;,
\label{pin1}
\end{equation}
where the integration over $\lambda$ ensures that
\begin{equation}
S\;=\;-\,\sum_{m = m_1+1}^{m = m_2} s_m 
\label{S}
\end{equation}
and the limits of the integration over $S$ follow from the first 
$\Theta$-function in Eq.~(\ref{pin}).

Since the typical pair size $R^{\ast} \gg 1$ [see Eq.~(\ref{estpeak})],
we can use the canonical formalism, in which Eq.~(\ref{S}) for the
sum of $R$ random variables is satisfied only in average.
We do this by ``shifting'' the argument of the random-field distribution 
on each site by $\lambda$:
\begin{equation}
f(s) \rightarrow f(s+\lambda) = e^{-\frac{1}{2}\lambda^2 - \lambda s} f(s)\;,
\end{equation}
so that the average value now becomes $s = -\lambda$ and
Eq.~(\ref{pin1}) reads:
\begin{equation}
p_{in} = \int_{zR+I}^{\infty}dS
\int_{-i\infty}^{+i\infty}\!\!\frac{d\lambda}{2\pi i}
e^{-\lambda S+ \frac{1}{2}R\lambda^2}\prod_{m=m_1+1}^{m_2}
\int_{-\infty}^{+\infty}\!\!ds_m f(s_m+\lambda) \Pi_L \Pi_R\;.
\label{pin1.1}
\end{equation}
The integral over $\lambda$ comes from the small vicinity 
($\sim 1/\sqrt{R}$) of
${\lambda_0} = S / R$, where the exponential in Eq.~(\ref{pin1.1}) has
its maximum. Saddle-point integration over $\lambda$ then gives:
\begin{equation}
p_{in} = \int_{zR+I}^{\infty}\frac{dS}{\sqrt{2 \pi R}}
e^{-\frac{S^2}{2R}}\prod_{m=m_1+1}^{m_2}
\int_{-\infty}^{+\infty}\!\!ds_m f\left(s_m+\frac{S}{R}\right)
\Pi_L \Pi_R\;.
\label{pin2}
\end{equation}

Next we note that if the condition imposed by the first
$\Theta$-function in Eq.~(\ref{pin}) is satisfied, the
arguments of the last $\Theta$-functions in $\Pi_L$ and $\Pi_R$
also almost certainly are positive. 
In other words, because the relevant disorder realizations are close
to the optimal fluctuations, only the first few $\Theta$-functions in 
$\Pi_L$ and $\Pi_R$ are really restrictive. 
This implies that the disorder averages of
$\Pi_L$ and  $\Pi_R$ in Eq.~(\ref{pin2}) are decoupled.
Furthermore, it is easily seen from Eq.~(\ref{yzla}) that then
$\langle\Pi_L\rangle=\langle\Pi_R\rangle=Y(\frac{S}{R}-z)$, 
so that Eq.~(\ref{pin2}) becomes:
\begin{equation}
p_{in} = \int_{zR+I}^{\infty}\frac{dS}{\sqrt{2 \pi R}}
e^{-\frac{1}{2R}S^2}
\left[Y\left(\frac{S}{R}-z\right)\right]^2\;.
\label{pin3}
\end{equation}
The integral over $S$ comes from the vicinity of the
lower limit, $S = zR+I$ [cf. Eq.~(\ref{ebal}) for the optimal
fluctuation]. The result of the integration is:
\begin{equation}
p_{in} = \sqrt{\frac{R}{2\pi}}\,
\frac{\exp\left[- \frac{(I + Rz)^2}{2R}\right]}{(I + Rz)}
\left[Y\left(\frac{I}{R}\right)\right]^2\;,
\label{pin4}
\end{equation}
where for $S$ in the argument of $Y$ we took its value at the
lower limit of the integration.

From Eqs.~(\ref{pfactorize}), (\ref{pout1}), and (\ref{pin4}) 
we finally obtain for the pair size distribution:
\begin{equation}
p(R)\;=\;
\frac{\exp(-g(R)^2)}{\sqrt{4\pi}\:g(R)}\;
\left[Y\left(\frac{I}{R}\right)\right]^2\,\left[Y(z)\right]^2\;,
\label{sortofsol}
\end{equation}
where the function $g(R)$ is defined by Eq.~(\ref{defofg}).

What is left now, is the calculation of the function $Y(v)$. To
this end we introduce the function $Y(s|v)$, satisfying
the integral equation:
\begin{equation}
Y(s|v)\;=\;
\int_{0}^{\infty}ds^{\prime}\,f(s+v-s^{\prime})\:
Y(s^\prime|v)\;.
\label{iteraY}
\end{equation}
Comparing the iterative solution of this equation to Eq.~(\ref{yzla}),
one finds:
\begin{equation}
Y(v)\;=\;Y(0|v)\;.
\label{szerolimit}
\end{equation}
The integral equation (\ref{iteraY}) can be easily solved
numerically.  The result is shown as stars in Fig.~2.  The solid
line represents the best fit to these points by a function of the
form:
\begin{equation}
Y(v)\;=\;\tanh(c\,v)\;.
\label{Yfit}
\end{equation}
The fit yields $c \simeq 1.14$.
For small $v$, the best linear fit $(Y(v)=c^{\prime}v)$ yields
$c^{\prime}\simeq\sqrt{2}$, with a precision of several percent.

We conclude this section by a brief analysis of the pair size
distribution Eq.~(\ref{sortofsol}).
First, integrating $p(R)/d$ over $R$ gives the density of 
soliton-antisoliton pairs [cf. Eq.~(\ref{denspofR})].  
For $\epsilon \ll \epsilon_c$, the exponential factor in 
Eq.~(\ref{sortofsol}) has a sharp peak at $R^{\ast}$ given by 
Eq.~(\ref{estpeak}).  
Using $B\simeq W$, the saddle-point integration around the peak 
gives:
\begin{equation}
n_p\;\approx\;
\frac{1}{d}\,\frac{\epsilon}{2\,W^2}\,
\left[Y\!\left(\frac{W}{\sqrt{\epsilon}}\right)\right]^{\! 4}
\exp\left(-2\frac{W\mu}{\epsilon}
\right)\;.
\label{densdiscres}
\end{equation}
Furthermore, we note that knowledge of the pair size distribution 
Eq.~(\ref{sortofsol}) allows us to derive the long- and short-range 
bond order parameter of the RFIM Eq.~(\ref{rfim}).
For a dilute gas of soliton-antisoliton pairs it is sufficient to 
consider a chain segment of $N+1$ sites containing a single pair 
of size $R$:
\begin{equation}
\sigma(m)\;=\;1\,-\,2\,\Theta(m-m_1)\,\Theta(m_2-m)\;,
\label{contsig}
\end{equation}
where $m_1$ and $m_2=m_1+R$ denote the positions of, respectively,
the soliton and the antisoliton.
Replacing summations by integrations, the LRBO parameter averaged
over all possible pair sizes is easily calculated: 
\begin{equation}
\langle\langle\sigma\rangle\rangle\;=\;\int_{0}^{\infty}\!\!dR\,p(R)\,
\int_{-\frac{N}{2}}^{+\frac{N}{2}}\!dm\,\sigma(m)\;=\;
1\,-\,2\,\int_{0}^{\infty}\!\!dR\,p(R)\,R\;.
\label{lrboparaexp}
\end{equation}
For a nearly perfectly ordered system, 
$\langle\langle\sigma\rangle\rangle\simeq 1$, we thus find from 
Eq.~(\ref{lrboparaexp}) the condition 
that the typical pair size $R^{\ast}$ is much smaller than the typical 
number of sites $1/(d n_p)$ between soliton-antisoliton pairs.
Similarly, we calculate the correlation function 
$\langle\langle\sigma(0)\sigma(l)\rangle\rangle$,
which yields the sum of the square of the LRBO parameter 
Eq.~(\ref{lrboparaexp}), 
\begin{equation}
\langle\langle\sigma\rangle\rangle^{2}\;\simeq\;
1\,-\,4\,\int_{0}^{\infty}\!\!dR\,p(R)\,R\;,
\end{equation}
and the connected correlator 
\begin{equation}
\langle\langle\sigma(0)\sigma(l)\rangle\rangle_c\;=\;
4\,\int_{0}^{\infty}\!\!dR\,p(R+|l|)\,R\;.
\label{conneccorrel}
\end{equation}
The scale for the decay of short-range correlations is obviously set 
by the typical pair size $R^{\ast}$, as 
$\langle\langle\sigma(0)\sigma(l)\rangle\rangle_c\rightarrow 0$ for 
$|l|\gg R^{\ast}$.
%
%
%

\section{The Exchange Distribution in the Continuum Limit}
\label{conti}

%
%
%
In the previous sections we described disordered Peierls systems
using the effective RFIM Eq.~(\ref{rfim}). The values of the
interchain interaction $W$ and the disorder strength $\epsilon$
in this model are proportional to the length $d$ which we choose
for the unit cell of the Ising chain and which plays the role of 
a short-distance cut off.
On the other hand physical observables, such as the density
of soliton-antisoliton pairs and the magnetic susceptibility,
should not depend on $d$. 
Thus, before comparing our results to the experimental data on 
{\it trans}-polyacetylene, we show that $d$ drops out from the 
expressions for the observables in the $d\rightarrow 0$ limit.

To this end, we introduce as physically meaningful quantities the
disorder strength $\bar{\epsilon}$ and the interchain interaction
$\bar{W}$ per unit length:
\begin{equation}
\epsilon\;=\;{\bar \epsilon}\,d\;,
\end{equation}
and
\begin{equation}
W\;=\;{\bar W}\,d\;.
\label{Wbar}
\end{equation}
Furthermore, from now on we will work with the physical pair size
$r = R d$.
In terms of these new variables, the arguments of both $Y$-functions in
Eq.~(\ref{sortofsol}) for the pair size distribution
are $\propto \sqrt{d}$. 
Therefore, for $d \rightarrow 0$, the arguments are small and we can 
use $Y(v)\simeq \sqrt{2} v$ [see below Eq.~(\ref{Yfit})], giving the pair 
size distribution
\begin{equation}
\bar{p}(r)\;=\;
4\,
\frac{\mu^2\,\bar{W}^2}{\bar{\epsilon}^{\,2}\,r^2}\,
\frac{\exp\left(-g(r)^2\right)}{\sqrt{4\pi}\:g(r)}\;,
\label{pofRbar}
\end{equation}
where
\begin{equation}
g(r)\;=\; \frac{\mu\,+\,\bar{W}\,r}
{\sqrt{2\,\bar{\epsilon}\,r}}\;.
\label{gofRbar}
\end{equation}
Similarly, from Eq.~(\ref{densdiscres}), the total density of 
neutral soliton-antisoliton pairs in the limit $d\rightarrow 0$
is found to be:
\begin{equation}
n_p
\;\approx\;
\frac{2\,\bar{W}^2}{\bar{\epsilon}}\,
\exp\left(-2\frac{\bar{W} \mu}{\bar{\epsilon}}\right)\;,
\label{densresfin}
\end{equation}
which coincides with Eq.~(\ref{contidens}) obtained in Ref.~\cite{Figge98} 
within the continuum approximation for the RFIM Eq.~(\ref{rfim}).

Furthermore, in terms of continuum variables, the exchange coupling
Eq.~(\ref{JofRandrho}) reads:
\begin{equation}
J(r)\;=\;J_0\,\exp(-\frac{r}{\xi_0})\;.
\label{JofRbar}
\end{equation}
Thus, in the continuum limit the distribution of exchange constants 
becomes:
\begin{equation}
w(J)
\;=\;\int_0^{\infty}\!\!dr\,\bar{p}(r)\,
\delta\!\left(J-J(r)\right)
\;=\;
4\,\frac{1}{J}\,\frac{1}{\xi_0}
\frac{\mu^2\,\bar{W}^2}{\bar{\epsilon}^{\,2}\,[\ln(J/J_0)]^2}
\,\frac{\exp\left(-g(J)^2\right)}{\sqrt{4\pi}\:g(J)}
\;,
\label{randexconti}
\end{equation}
with $g(J)\equiv g(r=\xi_0\ln(J_0/J))$ 
[cf. Eq.~(\ref{gofRbar})].

In Fig.~3 we plot the distribution $w(J)$ for four different
parameter sets $(\bar{\epsilon}, \bar{W}, J_0)$ chosen such that the
density of soliton-antisoliton pairs is fixed at $n_p=1/6000\,a^{-1}$
(with $a$ the average carbon-carbon distance in {\it trans}-polyacetylene).
Our choice of parameters is summarized in Table~I and will become clear 
in Sec.~\ref{appl}.
Depending on the parameters, one observes two qualitatively very different
behaviors: $w(J)$ either has a pronounced peak at $J^{\ast}\simeq
J(r^{\ast})$ (with $r^{\ast}=R^{\ast}\,d$ and $R^{\ast}$ as in
Eq.~(\ref{estpeak})) and tends to zero for $J\rightarrow 0$, or $w(J)$ diverges 
for small $J$.
The distinction between these two behaviors is dominated by only one
parameter combination:
\begin{equation}
\alpha\;=\;\frac{\bar{W}^2}{2\,\bar{\epsilon}}\,\xi_0\;.
\label{polyalpha}
\end{equation}
In fact, Eq.~(\ref{randexconti}) for $J\ll J^{\ast}$ yields
\begin{equation}
w(J)\;\propto\;\left(\frac{J_0}{J}\right)^{1-\alpha}\;,
\label{wJsmallJ}
\end{equation}
which shows that the relative strength of the interchain interactions
and the disorder determines whether $w(J)$ diverges ($\alpha<1$)
or approaches zero ($\alpha>1$) for $J\rightarrow 0$.

The behavior of Eq.~(\ref{wJsmallJ}) agrees with Eq.~(\ref{powerlawex})
in Sec.~\ref{masu} and can indeed be traced back to the fact that for large
$r$ the pair size distribution is exponential:
\begin{equation}
\bar{p}(r\gg r^{\ast})\;\propto\;
\exp\left(-\frac{\bar{W}^2\,r}{2\bar{\epsilon}}\right)\;=\; 
\left(\frac{J_0}{J}\right)^{-\alpha}\;.
\label{qualexpla}
\end{equation}
This exponential dependence can be understood as follows. 
For $r\gg r^{\ast}$, the energy of the string between soliton
and antisoliton exceeds the kink creation energy: $\bar{W}r=WR\gg \mu$.
Thus, the amplitude of the optimal fluctuation Eq.~(\ref{ebal}) is
$h_R R \simeq W R$.
The Gaussian weight $\exp\left(-R h^2_R /(2\epsilon)\right)$ of this
fluctuation is the exponential in Eq.~(\ref{qualexpla}).
Similar arguments were used to explain the power-law dependence of
the density of states in the Fluctuating Gap Model of disordered systems
\cite{Fluct}.
The power-law dependence of $w(J)$ at small $J$ gives rise to a
characteristic low-temperature behavior of the magnetic susceptibility,
as we will see in the next section.
%
%

\section{Magnetic susceptibility of {\it trans}-polyacetylene}
\label{appl}

%
%
In this section we consider {\it trans}-polyacetylene as a
disordered Peierls system and calculate its magnetic susceptibility 
due to disorder-induced soliton-antisoliton pairs as a function of
temperature.
The temperature dependence of the magnetic susceptibility is determined
by the distribution of exchange constants Eq.~(\ref{randexconti}). 
For temperatures $T$ much larger than the typical singlet-triplet energy
splitting $J^{\ast}$, almost all spin pairs are thermally excited.
Thus, we have, essentially, $2n_p$ free spins, which give
rise to a Curie susceptibility. 
Indeed, Eqs.~(\ref{pairsus}) and (\ref{chires}) yield:
\begin{equation}
\chi(T\gg J^{\ast})\;\simeq\;
\frac{1}{2}\,g^2\mu_B^{2}\,\beta\,
\int_{0}^{\infty}\!\!dJ\,w(J)\;=\;
\frac{1}{2}\,g^2\mu_B^{2}\;
\frac{n_p}{T}\;,
\label{sushightcont}
\end{equation}
where the density of soliton-antisoliton pairs $n_p$ is given by 
Eq.~(\ref{densresfin}).
In the opposite limit, $T\ll J^{\ast}$, however, most of the spin
pairs are in the singlet state and do not contribute to the
magnetic susceptibility.
Under these conditions Eqs.~(\ref{chires}) and (\ref{randexconti})
yield:
\begin{equation}
\chi(T\ll J^{\ast})\;=\;
C(T)\,\left(\frac{J_0}{T}\right)^{\!1-\alpha}\;,
\label{suslowtcont}
\end{equation}
with $\alpha$ as in Eq.~(\ref{polyalpha}) and logarithmic temperature
corrections given by
\begin{equation}
C(T)\;\approx\;\frac{4\,g^2\,\mu^2_B}{9\,J_0\,\alpha}\,
\left(\frac{n_p\,\xi_0}{\pi}\right)^{\!\frac{1}{2}}\,
\frac{\mu^2}{\bar{\epsilon}\,\xi_0^2}
\,\frac{\Gamma(1+\alpha)\,{\rm Li}_\alpha(-\frac{1}{3})}{(\ln(J_0/T))^{5/2}}
\end{equation}
(${\rm Li}_{\nu}(z)=\sum_{k=1}^{\infty}z^k/k^\nu$ is the polylogarithm 
function).
The dominant factor in Eq.~(\ref{suslowtcont}) is $(J_0/T)^{1-\alpha}$,
which basically gives the density of spin pairs with singlet-triplet
splitting $\sim T$ [cf. Eq.~(\ref{wJsmallJ})].

We thus find that the low-temperature behavior of the magnetic
susceptibility differs from the Curie law and is dictated by the
relative strength $\alpha$ of interchain interactions and disorder.
For $\alpha<1$, the susceptibility diverges as $T\rightarrow 0$,
while for $\alpha>1$ it approaches zero.

A low-temperature ($T<30$ K) deviation from Curie behavior has indeed
been observed by Foot {\it et al.} in ESR experiments on Durham
{\it trans}-polyacetylene \cite{Foot86}.
These authors already suggested pairing of spins as possible reason
for this behavior.
To see whether our model of spins associated with disorder-induced
soliton-antisoliton pairs offers a microscopic explanation,
we compared our result for $\chi(T)$ [numerically calculated
from Eqs.~(\ref{chires}) and (\ref{randexconti})] to the experimental
data.
In our fit procedure there are, in principle, three free
parameters: $\bar{W}$, $\bar{\epsilon}$, and $J_0$ (for the correlation
length we take the SSH value $\xi_0=7a$).
We require, however, that our parameters are also consistent with
the total density of spin pairs, which is reported to be
approximately $n_p=1/6000\,a^{-1}$ 
\cite{Shirakawa78,Goldberg79,Weinberger80,Foot86}.
This requirement imposes a relation between $\bar{W}$ and $\bar{\epsilon}$,
reducing the number of free parameters to two.

In practice, we chose various values for $\alpha$ [Eq.~(\ref{polyalpha})].
For each $\alpha$ value, $\bar{W}$ and $\bar{\epsilon}$ are uniquely
determined by $n_p$, and $J_0$ is left as free parameter to fit
the temperature dependence of the magnetic susceptibility.
This procedure yielded the fits shown in Fig.~4, with parameter sets
given in Table I.
For convenience, we will refer to each parameter set by its $\alpha$
value.
We note that our values for $J_0$ are much smaller than the value
$J_0=4\Delta_0\sim 10^{4}$ K, which one would expect in the absence
of Coulomb interactions.
It is not known, however, how Coulomb interactions alter $J_0$ and
(possibly) the exponent in Eq.~(\ref{JofRandrho}).

The important point is now that, while all four parameter sets give
rise to reasonable fits of the experimental data, they predict totally
different behaviors for $T\leq 5$ K, where experiments have not been
performed. 
This is shown in Fig.~5, where we extend the four theoretical
fits to $1$ K.
The qualitative differences in the low-temperature behavior, dictated by
the value of $\alpha$, are clearly visible below $5$ K.
This suggests that extending the experiments to lower temperatures
may yield more information on the relative strength of interchain 
interactions and disorder in {\it trans}-polyacetylene.

We conclude this section by noting that, within the context of our model,
it is possible to determine the strength of the interchain interactions and 
the disorder independent of the fitting parameter $J_0$.
For this purpose, the density $n_p$ of soliton-antisoliton pairs is to be 
obtained experimentally from the Curie tail of the magnetic susceptibility,
while for the same sample $\alpha$ is to be determined from the asymptotic 
zero-temperature behavior of the magnetic susceptibility.
Then, using Eqs.~(\ref{densresfin}) and (\ref{polyalpha}), the strength 
of the interchain interactions,
\begin{equation}
\bar{W}\;\approx\;4\,\alpha\,\frac{\mu}{\xi_0}\,
\left[\ln\left(\frac{4\alpha}{n_p\xi_0}\right)\right]^{-1}\;,
\label{Wwellwithin}
\end{equation}
and the disorder strength,
\begin{equation}
\bar{\epsilon}\;\approx\;8\,\alpha\,\frac{\mu^2}{\xi_0}\,
\left[\ln\left(\frac{4\alpha}{n_p\xi_0}\right)\right]^{-2}\;,
\label{epswellwithin}
\end{equation}
can be calculated as a function of $n_p$ and $\alpha$.
For typical values of the density $n_p$ the logarithmic factor 
depends only weakly on $\alpha\sim O(1)$ and can be approximated by 
a numerical constant.
If $n_p=1/6000\,a^{-1}$ and choosing again the SSH-parameter $\xi_0= 7a$, 
one obtains $\bar{W}\,\approx\,0.07\,\alpha\,\mu/a$ and 
$\bar{\epsilon}\,\approx\,0.02\,\alpha\,\mu^2/a$.
It is important to realize that Eqs.~(\ref{Wwellwithin}) 
and (\ref{epswellwithin}) do not depend on the maximal exchange 
constant $J_0$ which may be used as a fitting parameter for temperatures 
around the typical exchange $J^{\ast}$.
%
%

\section{Concluding remarks}
\label{sumconcl}

%
%
To summarize, we have calculated the magnetic susceptibility of
quasi-one-dimensional Peierls systems with a 
doubly degenerate ground state.  We have related the
temperature-dependent part of the susceptibility to the presence
of neutral solitons and antisolitons with spin $1 \over 2$
induced by disorder in the electron hopping amplitudes along the chain.  We
have assumed the interchain interactions to be sufficiently strong
to bind the disorder-induced solitons and antisolitons into
pairs and thus establish long range bond order in the system.
Using a mapping on the random field Ising model, we have calculated the
distribution of the soliton-antisoliton pair size. This
allowed us to obtain the distribution of exchange constants
describing the interaction between the spins of the soliton and the
antisoliton within one pair.  Both distributions strongly depend on
the relative strength $\alpha$ of the disorder and the interchain
interactions.  As a result, the magnetic susceptibility deviates
from the Curie law: below $T = J^{\ast}$, where $J^{\ast}$ is the
most probable value of the exchange constant, the magnetic
susceptibility behaves as $(1/T)^{1-\alpha}$.

Our results explain the deviation from Curie behavior
observed in Durham {\it trans}-polyacetylene \cite{Foot86},
though from the experimental data it is difficult to find
unambigously the values of $\alpha$ and $J^{\ast}$ for this
conjugated polymer.  It is, therefore, important to extend the
measurements to lower temperatures, where the temperature
dependence of the magnetic susceptibility is extremely sensitive
to the choice of parameters.

Our theory is only applicable when the low-temperature behavior of
the susceptibility is an intrinsic property of the material and
is not governed by spins of impurities.  The latter situation may, in fact,
be realized in Shirakawa {\it trans}-polyacetylene, which
shows Curie behavior down to $T=1.5$ K \cite{Schwoerer80}.
Furthermore, we assumed the existence of long range order in
the system.  Whether this is the case in {\it trans}-polyacetylene is an
open question.  It would, therefore, be interesting to extend our studies
to the case without long-range order.  At the same time, however, it should
be noted that in substituted polyacetylenes, the degeneracy of the two
dimerized configurations may be lifted.  This leads to an extra
(intrachain) source of soliton-antisoliton confinement\cite{Kobayashi96} 
and favors long-range bond order.  Our theory may be applied to these
substituted polymers by simply adding to the interchain interaction per
bond ($Wa$), the energy difference per bond between the two dimerized
configurations.  As this energy difference may be controlled by varying the
substitutions, this opens interesting possibilities to study
disorder-induced solitons in more detail.  

We finally mention that in the ordered phase, the disorder-induced
soliton-antisoliton pairs show up in the x-ray spectrum as a broad
incoherent peak associated with each sharp elastic peak arising from the
bond length alternation.  Our result for the pair size distribution allows
one to calculate the shape of this incoherent peak: it simply is the
Fourier transform squared of the connected correlator 
Eq.(\ref{conneccorrel}).  
Thus, one immediately finds that the peak width is 
$\sim 1/r^{\ast} \approx \bar{W}/\mu$.  
It should be kept in mind, however, that this calculation does
not account for other broadening mechanisms, e.g., those due to the
complicated morphology of polyacetylene samples.  
%
%
%
\section*{Acknowledgments}
%
%
%
This work is part of the research program of the Stichting
Fundamenteel Onderzoek der Materie (FOM), which is financially
supported by the Nederlandse Organisatie voor Wetenschappelijk
Onderzoek (NWO).
%
%
%

%
%
%
%
%
\newpage
\begin{center}
{\bf Tables}
\end{center}
\vspace{1cm}
%
%
\begin{tabular}{|c|c|c||c|} \hline
$\;\;\;\bar{W}\;\,(\mu/a)\;\;\;$ & 
$\;\;\;\bar{\epsilon}\;\,(\mu^2/a)\;\;\;$ &
$\;\;\;J_0\;\,(K)\;\;\;$ & 
$\;\;\;\;\alpha\;\;\;\;$ 
\\ \hline\hline
0.038 & 0.010 & 400 & 0.5 \\ \hline
0.055 & 0.014 & 130 & 0.75 \\ \hline
0.070 & 0.017 & 72 & 1.0 \\ \hline
0.100 & 0.023 & 39 & 1.5 \\ \hline
\end{tabular}%

\vspace{1.0cm}

\noindent Table~I. The four sets of parameters used in the 
numerical calculations for a fixed density of spin pairs, 
$n_p=1/6000\,a^{-1}$.
The value for $\alpha$ as defined in Eq.~(\ref{polyalpha}) 
is obtained using the SSH-parameter $\xi_0=7a$ for the 
correlation length.  
\newpage
%
%
\begin{center}
{\bf Figures}
\end{center}
\vspace{1cm}
\begin{center}
\includegraphics[width=12cm]{fig1.epsi}
\end{center}
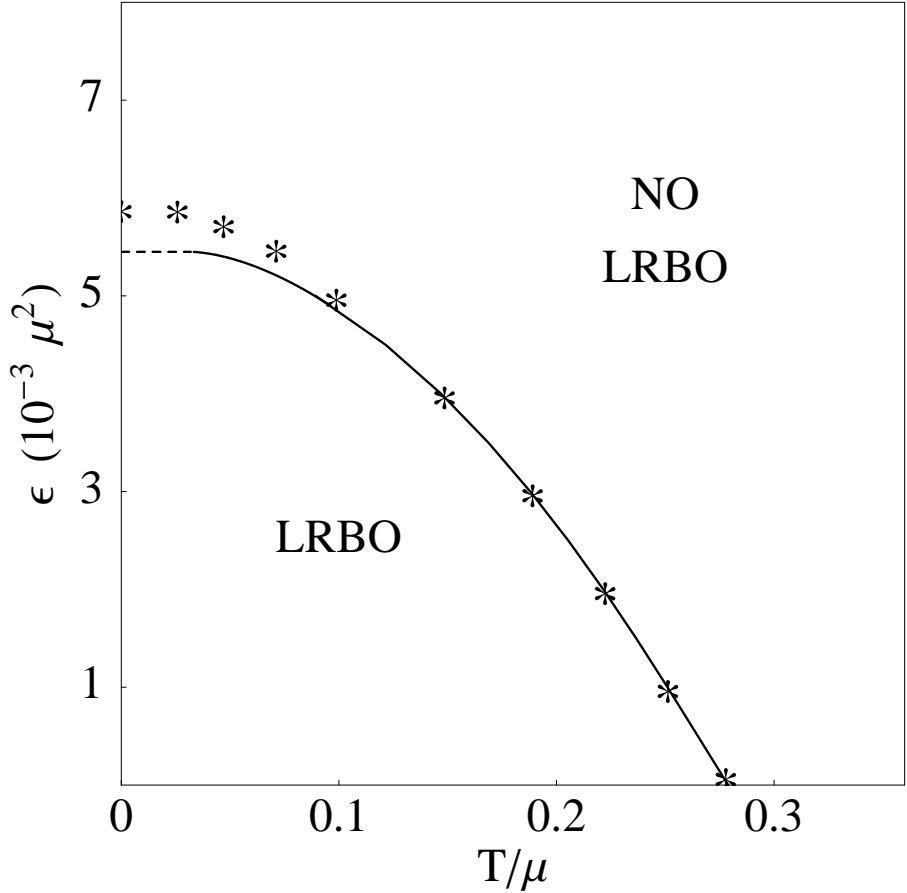
Fig.~1.  The phase diagram of the RFIM Eq.~(\ref{rfim}) 
captures the essential physics of weakly disordered Peierls 
systems and is shown as a function of the disorder strength 
$\epsilon/\mu^2$ and the temperature $T/\mu$.
The long range bond order (LRBO) phase corresponds to an average
dimerization $\langle\langle\sigma\rangle\rangle\Delta_0\neq0$.
The numerical calculation of the critical curve (stars) agrees 
well with the analytical result (solid curve) obtained in 
Ref.~\cite{Figge98}. 
The dashed curve indicates the breakdown of the continuum 
approximation in the analytical calculation below $T=T_0(\epsilon)$ 
\cite{Figge98}.

\newpage

\begin{center}
\includegraphics[width=12cm]{fig2.epsi}
\end{center}
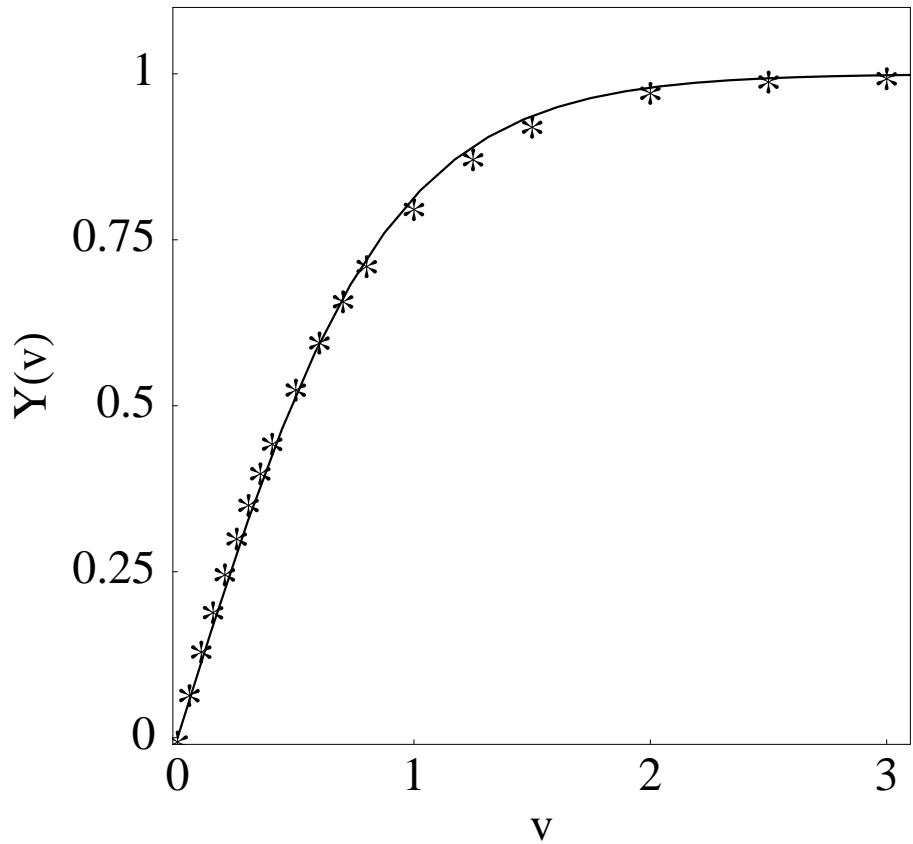
Fig.~2.  The function $Y(v)$ obtained from a numerical solution 
(stars) of the integral equation (\ref{iteraY}). The best fit of 
this solution by a function of the form $Y(v)=\tanh(cv)$ yields 
$c \simeq 1.14$ (solid line).    

\newpage

\begin{center}
\includegraphics[width=12cm]{fig3.epsi}
\end{center}
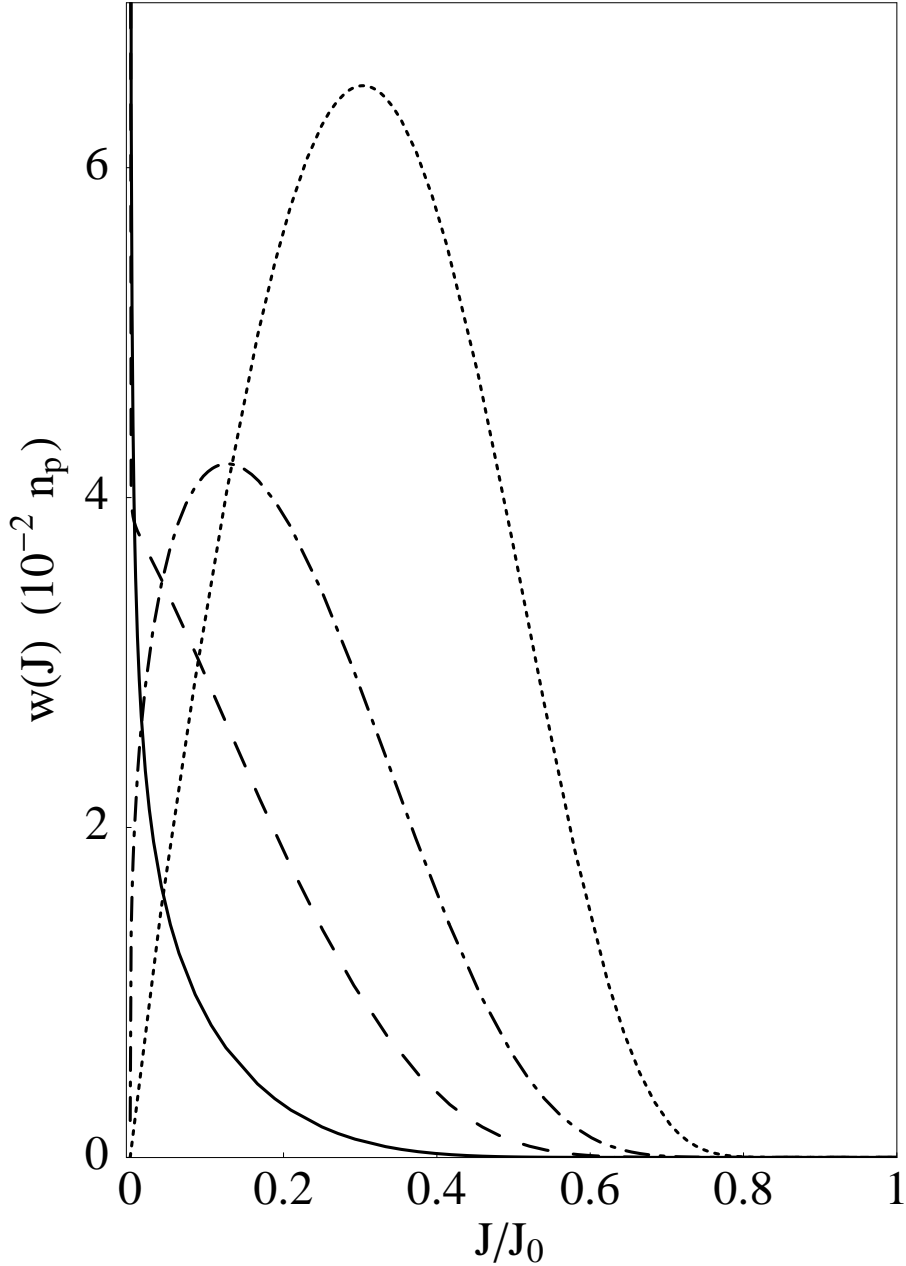
Fig.~3.  The distribution $w(J)$ of exchange constants as a function 
of $J/J_0$ for the four parameter sets given in Table~I. 
The curves correspond to $\alpha = 1.5$ (dots), $1.0$ (dash-dot), 
$0.75$ (dashes), and $0.5$ (solid).
For $\alpha\geq 1.0$ the distribution has a pronounced peak at some 
$J=J^{\ast}$ and tends to zero for $J\rightarrow 0$.
In contrast, for $\alpha < 1.0$, $w(J)$ diverges when $J\rightarrow 0$.

\newpage

\begin{center}
\includegraphics[width=12cm]{fig4.epsi}
\end{center}
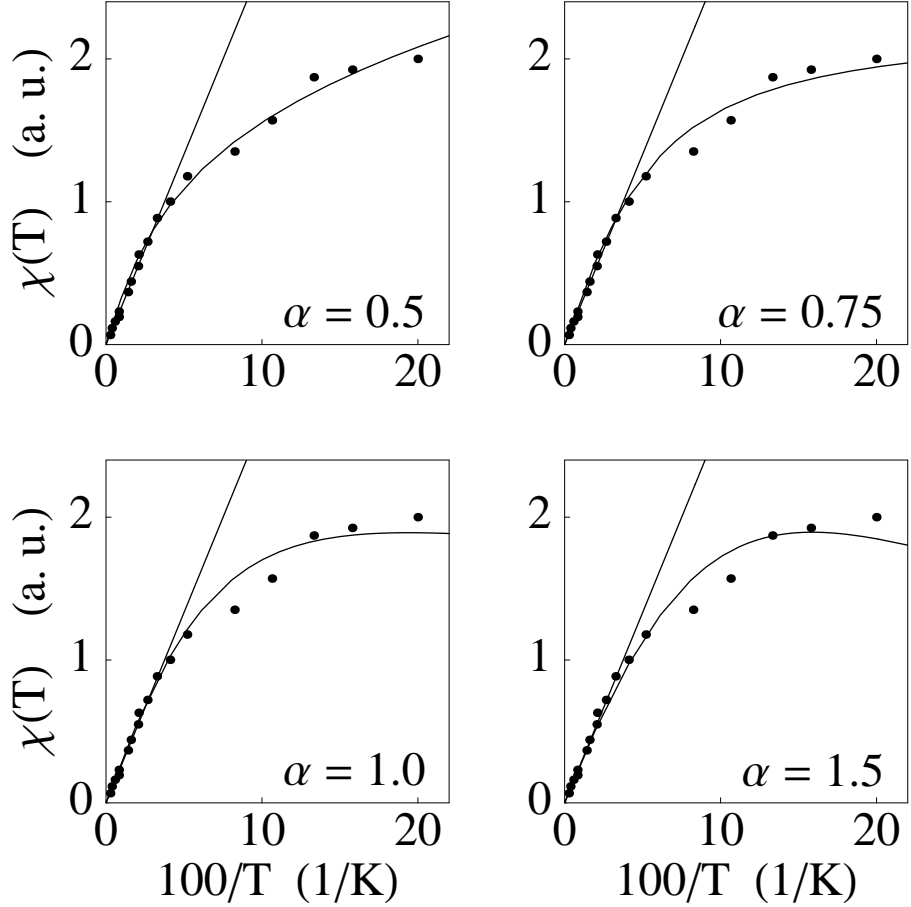
Fig.~4.  Fits of our theory (solid curve) to the experimental data 
(dots) for the magnetic susceptibility of Durham 
{\it trans}-polyacetylene obtained in Ref.~\cite{Foot86}. 
The four parameter sets given in Table~I were used to fit the same 
experimental data points (see text for details). 
The deviation from Curie behavior (straight line) below $T=30$ K is 
clearly seen and reasonably reproduced by each fit down to $T=5$ K, 
below which experimental data are not available. 

\newpage

\begin{center}
\includegraphics[width=12cm]{fig5.epsi}
\end{center}
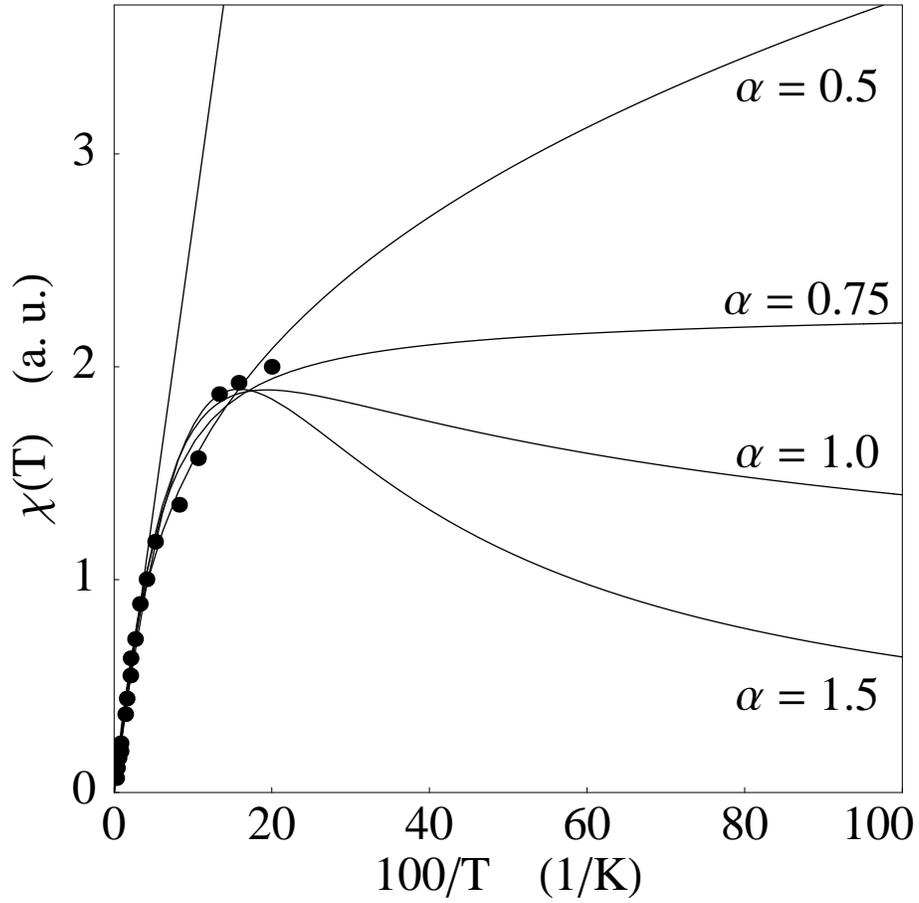
Fig.~5.  As Fig.~4, but now the four theoretical curves are shown 
down to $T= 1$ K.
It is clearly observed that different values for $\alpha$ (the 
relative strength of disorder and interchain interactions) lead to 
qualitatively different low-T ($< 5$ K) behavior of the magnetic 
susceptibility.
\end{document}